\documentclass[]{interact}

\usepackage{epstopdf}% To incorporate .eps illustrations using PDFLaTeX, etc.
\usepackage[caption=false]{subfig}% Support for small, `sub' figures and tables

\usepackage[numbers,sort&compress,merge]{natbib}% Citation support using natbib.sty
\bibpunct[, ]{[}{]}{,}{n}{,}{,}% Citation support using natbib.sty
\renewcommand\bibfont{\fontsize{10}{12}\selectfont}% Bibliography support using natbib.sty

\theoremstyle{plain}% Theorem-like structures provided by amsthm.sty

\usepackage[version=4]{mhchem} %chemical formula package
\usepackage{chemfig} %make chemical structure
\usepackage{gensymb} %make degree celsius
\usepackage{chemmacros}
\chemsetup{
  modules = {reactions} ,
  formula = mhchem
}%setup reaction number
\usepackage{physics}
\usepackage{stackengine}

\theoremstyle{definition}

\theoremstyle{remark}

\begin{document}

\articletype{ARTICLE TEMPLATE}% Specify the article type or omit as appropriate

\title{Chiral Pt(Me-BPCH): Synthesis and theoretical investigation of parity violation sensitivity $^\dag$}

\author{
\name{
Eduardus\textsuperscript{a},
J. Wietze J. van Boven\textsuperscript{a},
Charles Silva\textsuperscript{b},
Philip Karageorghis\textsuperscript{b},
D. Scott Bohle\textsuperscript{b},
Beno\^{i}t Darqui\'{e}\textsuperscript{c},
Anastasia Borschevsky\textsuperscript{a},
Lukáš F. Pašteka\textsuperscript{a,d}\thanks{CONTACT L.~F. Pašteka. Email: lukas.f.pasteka@uniba.sk}
}
\affil{
\textsuperscript{a}Van Swinderen Institute for Particle Physics and Gravity, University of Groningen, The Netherlands; 
\textsuperscript{b}Department of Chemistry, McGill University, Montréal, Canada;
\textsuperscript{c}Laboratoire de Physique des Lasers, CNRS, Universit\'{e} Sorbonne Paris Nord, Villetaneuse, France;
\textsuperscript{d}Department of Physical and Theoretical Chemistry, Comenius University,  Bratislava, Slovakia
}
}

\maketitle

\begin{abstract}
A complex of platinum and the tetra-coordinate chelating ligand, R,R'-6,6'-dimethyl-N,N’-bis(2’-pyridine-carboxamide)-1-cyclohexane (Me-BPCH) is investigated as a potential candidate for measurement of parity violation (PV) in chiral molecules.
The synthesis of Pt(Me-BPCH) is presented alongside computational investigation of PV sensitivity in its vibrational spectrum.
Pt(Me-BPCH) is compared to other two derivatives of this complex, Au(Me-BPCH) and Pt(CF$_3$-BPCH) in terms of their PV response and suitability for measurement.
We identify the most promising vibrational transitions based on their enhanced PV effects and practical experimental considerations and analyze the relationship between the vibrational structure and the corresponding PV sensitivity for all three molecules. 
\end{abstract}

%\begin{keywords}
%Sections; lists; figures; tables; mathematics; fonts; references; appendices
%\end{keywords}

\section{Introduction}

Since the discovery of molecular chirality by Louis Pasteur in 1848~\cite{pasteur}, this topic has captured the attention of the scientific world. Besides research on chirality induced spin selectivity (CISS)~\cite{CISS}, chiral molecule detection~\cite{Chiral_Detection}, and chiral molecule catalysis~\cite{chiral_catalysis}, the search for violation of parity (PV) in chiral molecules, that is, the search for a tiny difference between the energies of the two enantiomers of the same compound, has received significant theoretical and experimental attention~\cite{FieHaaSal22,Rhenium1,Rhenium2,chdbri1,chdbri2,Schwerdtfeger2010, Berger2019}.  
Parity violation has been successfully measured in nuclear and atomic experiments~\cite{Wu1,atomic1,atomic2}; however, so far only upper limits were set on the parity violating energy difference (PVED) in chiral molecules~\cite{CraChaSau05,DarStoZri10,Quack2022,DauMarAmy99,chardonnet:newexp}.
A precise measurement of this energy difference would constitute a stringent test of the Standard Model (SM) of particle physics in the low-energy regime \cite{safronova}, provide new insights into the nature of dark matter~\cite{gaul1}, and could shed new light on the origin of biohomochirality~\cite{Schwerdtfeger2010, Berger2019}.

Today, there are ongoing attempts to measure the  PVED in vibrational spectra of chiral molecules using high-precision lasers in the range of 500--2000 $\text{cm}^\text{--1}$~\cite{CouManPie19}. The success of these challenging experiments depends on the availability of strong intensity vibrational transitions that exhibit a  relative PV frequency shift ($\Delta \nu^\text{PV}/ \nu$) sensitivity at least in the order of $10^{-15}$, which is the projected sensitivity of state-of-the-art experiments~\cite{CouManPie19}. \textit{A priori} identification of such suitable candidate molecules requires computational investigations.

The PV energy scales as $\sim Z^5$~\cite{dovich,bast1,Bouchiat1974,Harris1978,Hegstrom1980} guiding the search for promising candidates towards heavy-metal containing compounds~\cite{dovich,bast1}. For example, a number of molecules containing period 6 heavy metals W, Re, Os, Ir, Au, Hg, Pb, Po and even actinide U were predicted to have large PV shifts in some of their vibrational transitions~\cite{Laerdahl19994439,Laerdahl20003811,Bast2003,Schwerdtfeger20031293,Schwerdtfeger20041652,Figgen2010,Figgen20102941,deMontigny20108792,Nahrwold2014,wormit2014strong,Rhenium1,Rhenium2}.

\begin{figure}
\begin{center}
\includegraphics[scale=0.40,angle=-90,trim={3cm 4cm 2cm 4cm},clip]{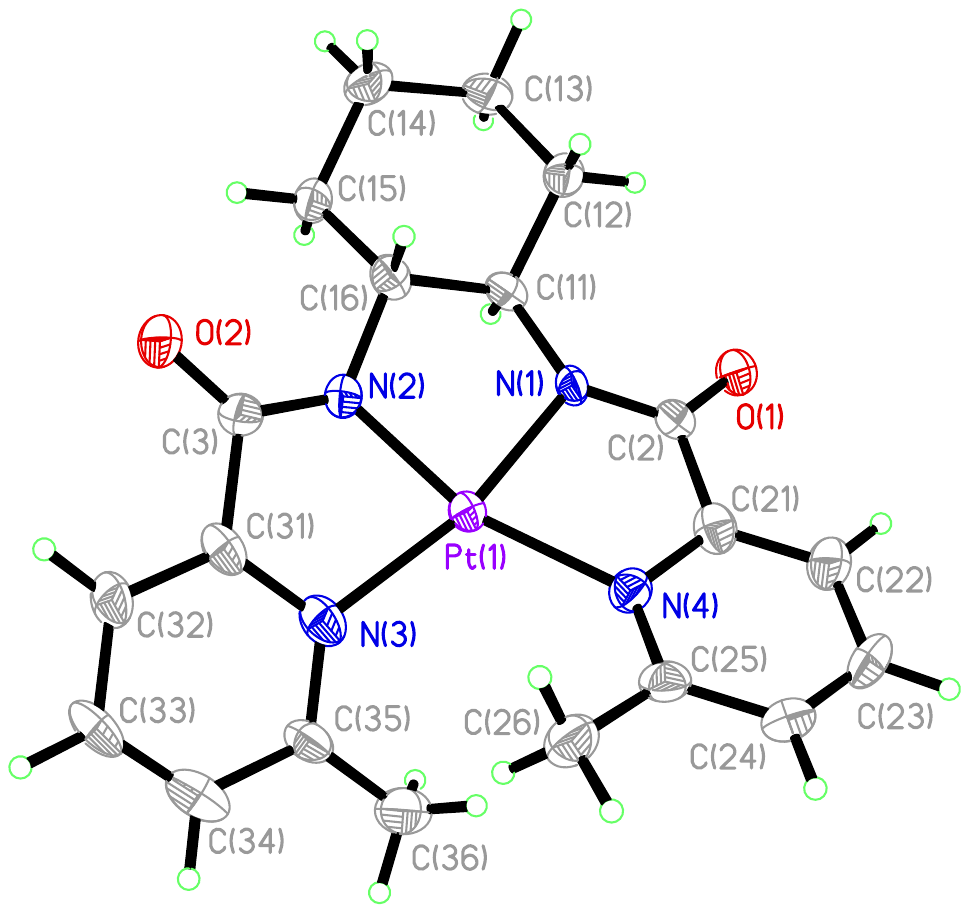}
\includegraphics[scale=0.30,angle=-90,trim={6cm 3cm 6cm 3cm},clip]{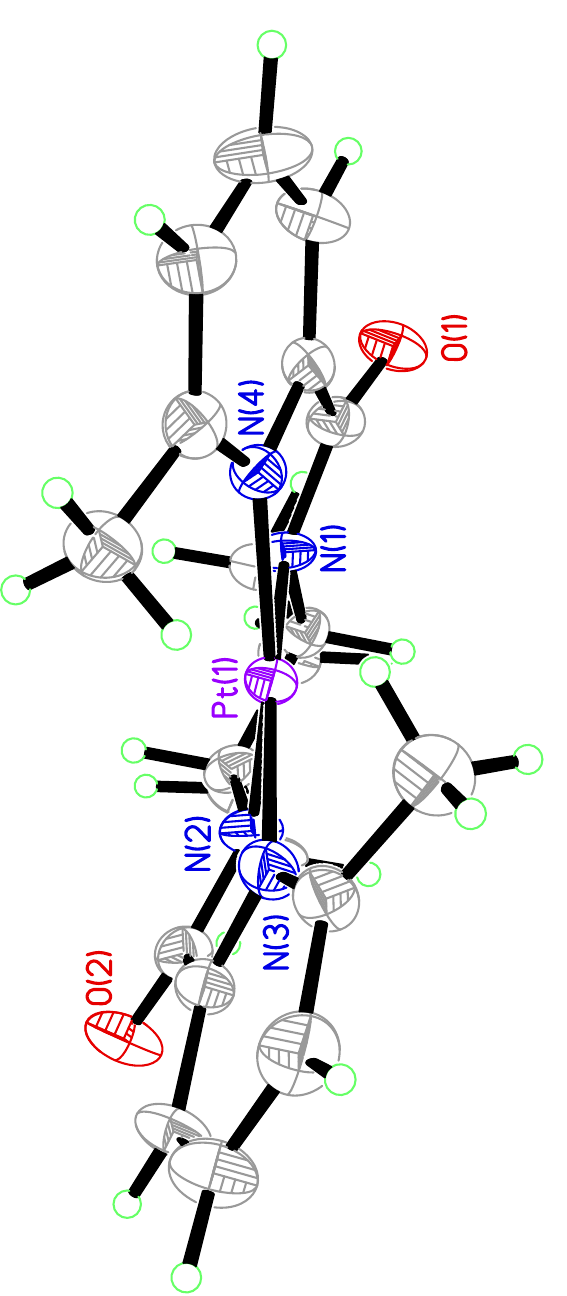}
\caption{Thermal ellipsoid plot of Pt(Me-BPCH). Note, the asymmetric unit contains two complexes.  Key metric parameters (\degree,\AA) include: Pt(1)-N(1) 1.971(8), Pt(1)-N(2) 1.954(7), Pt(1)-N(3) 2.064(3), Pt(1)-N(4) 2.069(7), N(1)-C(2) 1.327(12), C(2)-O(1) 1.251(11), N(2)-C(3) 1.343(11), C(3)-O(2) 1.224(11),  N(1)-Pt(1)-N(2) 85.2(3), N(3)-Pt(1)-N(4) 114.0(3). Top: view perpendicular to the C$_2$ axis; bottom: view along C$_2$ axis.}
\label{fig:structure}
\end{center}
\end{figure}

In this work, we focus on a candidate molecule with a C$_2$ axis of symmetry that runs through the central metal atom. Its chirality stems from the tetra-coordinate chelating ligand, R,R'-6,6'-dimethyl-N,N’-bis(2’-pyridine-carboxamide)-1-cyclohexane (Me-BPCH). The platinum complex Pt(Me-BPCH) is shown in Fig.~\ref{fig:structure}. We thus present the synthesis of this complex and computationally investigate its PV sensitivity by calculating the shifts in its vibrational spectrum resulting from these effects. 

In addition, we study its two  potential derivatives. In the first one, Pt is replaced by  Au(II) as the metal center, to obtain Au(Me-BPCH), which is aclosely related open-shell derivative of Pt(Me-BPCH). The interest on this molecule is motivated by our earlier finding that open shell systems can exhibit higher PV frequency shifts than their closed shell counterparts~\cite{chdbri1}).  Furthermore, the hexafluorinated version of Pt(Me-BPCH), Pt(CF$_3$-BPCH), is of considerable interest due to potentially higher volatility~\cite{Fcarbon}). Electronic structure calculations are used to predict the intensity of the different vibrational transitions and the size of the corresponding PV frequency shifts in the three molecules, with the goal to determine whether the expected effect is large enough to be detectable in a modern, state-of-the-art experiment~\cite{CouManPie19}. We identify the vibrational transitions that are the most promising for measurements, due to enhanced PV effects and because of practical considerations, such as wavelength that is easily accessible for commercial lasers and high transition intensity and analyze the relationship between the vibrational structure and the corresponding PV sensitivity for all three molecules.

\section{Material and Methods}

\subsection{Synthesis of Pt(Me-BPCH)}

The design of this candidate is based on the C$_2$ axial symmetry of the ligand which is derived from the amidation of R,R'-1,2-diaminocyclohexane (Reaction \ref{rea:1}). 
\begin{reaction}\label{rea:1}
\ce{\scalebox{0.9}{\chemfig{[,0.4]*6(-=-(-([2]=O)-[:-30]OH)=N-(-)=)}} ->[\stackon{\scalebox{0.8}{DMF (2 drops)}}{\scalebox{0.8}{\chemfig{[:-30,0.4]Cl-([6]=O)-[:30]([2]=O)-Cl}}\scalebox{0.8}{(1.8 eq.)}}][{DCM, 0{\degree}C, 5h}] {\scalebox{0.9}{\chemfig{[,0.4]*6(-=-(-([2]=O)-[:-30]Cl)=N-(-)=)}}} ->[\stackon{\scalebox{0.8}{\ce{Et3N} (3.66 eq.)}}{\scalebox{0.8}{\chemfig{[,0.4]*6(-(<NH_2)-(<:NH_2)----)}}\scalebox{0.8}{(0.5 eq.)}}][{DCM, 0{\degree}C, 4h}] {\chemname{\scalebox{0.8}{\chemfig{[,0.45]*6(=-(-)=N-(-([:150]=O)-[:30]NH(>*6(-(<:\chembelow{N}{H}-[:30]([2]=O)-*6(=N-(-)=-=-))-----)))=-)}}}{Me-BPCH}}}
\end{reaction}
This family of bis amidate chelating tetradentate ligands are known as good asymetric allylic alkylation catalysts~\cite{pflatz2004,belda2005,bateman2008}. Addition of the preformed ligand to Na$_2$[PtCl$_4$] returns a neutral diamagnetic tetradentate complex (Reaction \ref{rea:2}). 
\begin{reaction}\label{rea:2}
\ce{\ce{Me-BPCH} + \ce{Na2PtCl4} ->[\stackunder{\ce{H2O/EtOH}}{~\Delta, ~\text{2.5h}}][\text{--2HCl, --2NaCl}]\ce{Pt{(Me-BPCH)}}}
\end{reaction}
This initial product is purified to homogeneity by simple recrystallization. Crystals for X-ray diffraction (XRD) are obtained as a hydrate. 
The distorted square planar platinum geometry, Figure 1, results in a single diastereomer being produced in high yield, with facile access to its enantiomer being possible by starting with the S,S'-1,2-diaminocyclohexane. An important detail in the design of Pt(Me-BPCH) is the inclusion of methyl groups at the ortho positions on the pyridines.  These help accentuate the axial chirality of the chelate and complex.   To illustrate this axial twist we have calculated \cite{Mercury} the out-of-plane distortions for the two pyridine nitrogens from the plane defined by the platinum and the two amide nitrogens.  Thus, N(3) is 0.172~\AA~ above and N(4) is 0.336~\AA~ below the N(1)--Pt(1)--N(2) plane. Significantly, in spite of this sterically imposed helicity, the four metal bound nitrogen atoms retain a markedly square planar geometry, with the Pt(1) being less than 0.046~\AA~out of the plane defined by the four metal bound nitrogen atoms. Thus
the helicity steps from twisting to the two pyridine planes away from the square plane around
the platinum square plane. Full crystallographic, synthetic, and spectroscopic details, including alternative synthesis route of Au(Me-BPCH) and Pt(CF$_3$-BPCH) are contained in the Supplementary Material.

\subsection{Computational Details}

Geometry optimizations and the harmonic frequency analysis were carried out in Gaussian16 program~\cite{g16}, using the $\omega$B97X-D density functional~\cite{wb97xd} combined with the Def2-TZVPP basis set~\cite{def2_basis}, which includes scalar relativistic Stuttgart pseudopotential for the heavy metals (Pt, Au)~\cite{Def2_Pseudopotential}.

Subsequently, we calculated the PV frequency shifts for all the normal modes in the range of spectroscopic interest between 500--2000 $\text{cm}^\text{--1}$. For this investigation, we followed a similar procedure to that presented in Refs.~\cite{FieHaaSal22,chdbri1}. For each selected mode, energy calculations were performed along the normal mode coordinate $q$ at 11 evenly spaced points $q_i$ between --0.5 \AA~ and 0.5 \AA~ to obtain the potential energy curve $V(q)$, using the same level of theory as the geometry optimization. The total PV energy contributions were calculated for the same points along the normal mode in the relativistic program DIRAC~\cite{DIRAC23}, using the effective PV Hamiltonian as derived from the Z boson exchange between the electrons and nucleons~\cite{hpv_deriv}:
\begin{equation} \label{eq:HamiltonianPV}
\hat{H}_\text{PV}= \frac{G_\mathrm{F}}{2\sqrt{2}} \sum_A^\text{nuclei} Q_\mathrm{w}(A) \sum_i^\text{electrons} \gamma_i^5 \rho_A(\vec{R}_A-\vec{r}_i),
\end{equation} 
where $G_\mathrm{F}=2.22255\times 10^{-14}$ a.u. is the Fermi constant, $Q_\mathrm{w}(A)=(1-4\sin^2\theta_\mathrm{W})Z_A-N_A$ is the weak charge of nucleus $A$ with the Weinberg mixing angle $\theta_\mathrm{W}$ $(\sin^2\theta_\mathrm{W}=0.2312)$, and $Z_A$ and $N_A$ are the numbers of protons and neutrons inside nucleus $A$, respectively. $\gamma^5_i$ is the electron Dirac $\gamma^5$-matrix in the standard representation ($\gamma^5=i\gamma^0\gamma^1\gamma^2\gamma^3$) and $\rho_A(\vec{r})$ the nuclear charge density. 

\begin{figure}
\begin{center}
\includegraphics[scale=0.4]{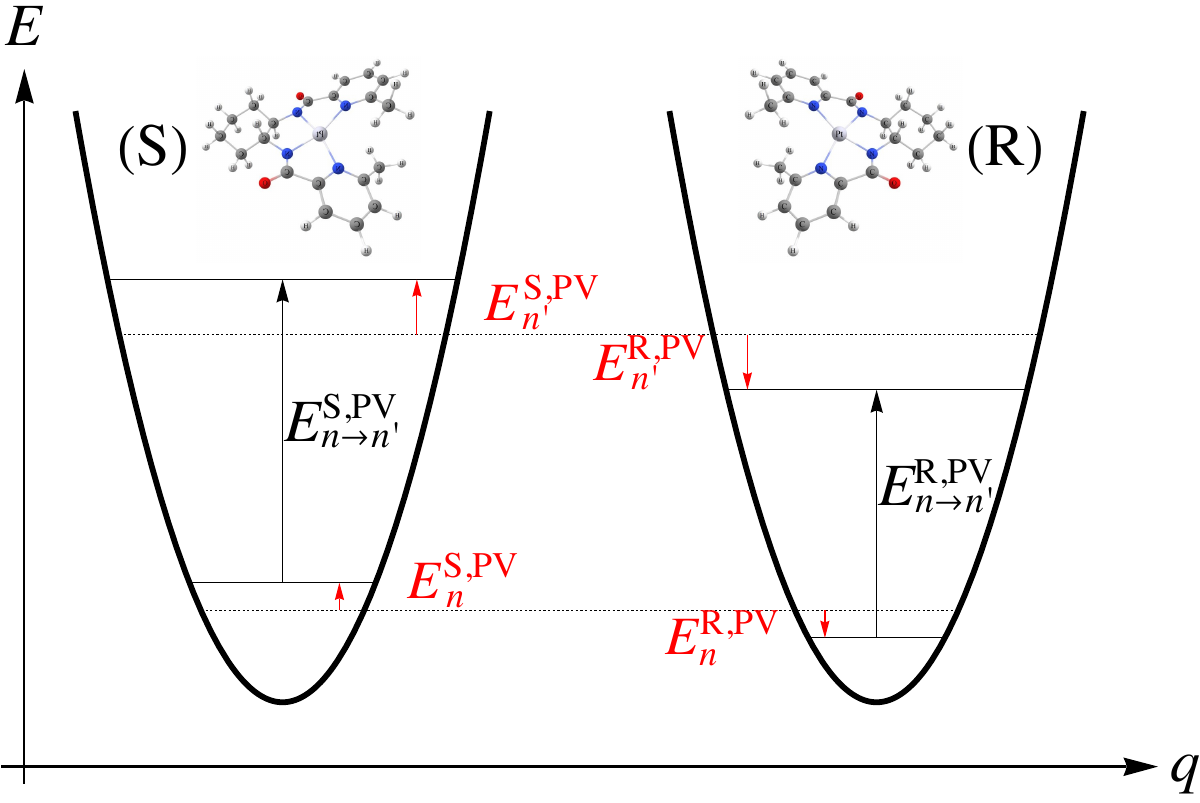}
\caption{Schematic representation of the left- (S) and right-handed (R) enantiomers of  Pt(Me-BPCH) and response of their respective vibrational transitions to parity violation effects.}
\label{fig:RS_well}
\end{center}
\end{figure}

The PV energy contribution at a given point along the normal mode coordinate $q$ is given by the expectation value  $V^\text{PV}=\bra{\Psi^\text{el}}H^\text{PV}\ket{\Psi^\text{el}}$. The $V^\text{PV}$ values are obtained using relativistic density functional calculations within the X2C/AMFI~\cite{Hess1996,AMFI} Hamiltonian and the CAM-B3LYP* functional optimized for PV calculations~\cite{ThiRauSch10}, combined with the uncontracted relativistic dyall.v3z~\cite{dyallRuv3z} basis set for the metal centers (Pt or Au) and nitrogens and dyall.v2z basis sets~\cite{dyallHv2z} for the rest of the atoms. 

To generate the PV shift for a specific vibrational mode $q$ we used the numerical Numerov--Cooley (NC) procedure~\cite{numerov1,Cooley_1961,Bast_2017}.
Both the potential  energy $V(q)$ and the PV energy contribution $V^\text{PV}(q)$ for the left handed enantiomer were fitted to a polynomial up to the sixth order. 
The vibrational wave functions $\ket{n}$ were determined by solving the vibrational Schr\"{o}dinger equation numerically. The parity violating energy shift of the $n$-th vibrational level was then calculated as $E^{PV}_n=\bra{n}V^\text{PV}(q)\ket{n}$.

%\begin{equation} \label{eq:E_pv_int}
%  E^{PV}_n=\bra{n}V_{PV}(q)\ket{n}.
%\end{equation}

The PV energy difference between the vibrational transitions of the two enantiomers (S,~R) is given by $\Delta E^\text{PV}_{n\rightarrow n'}=(E_{n\rightarrow n'}^\text{R,PV}-E_{n\rightarrow n'}^\text{S,PV})=2E_{n\rightarrow n'}^\text{R,PV}$ as illustrated in Figure \ref{fig:RS_well}. This energy difference corresponds to an absolute frequency shif  $\Delta\nu^\text{PV}_{n\rightarrow n'}=\Delta E^\text{PV}_{n\rightarrow n'}/h$. Finally, we obtain the relative frequency shift $\Delta \nu^\text{PV}_{n\rightarrow n'}/ \nu$, which is an important experimental predictor.

\subsection{Outline of the planned PV measurement}

At Laboratoire de Physique des Lasers (LPL), we aim to perform the first ever measurement of a PV vibrational frequency difference on gases of purpose-made chiral species cooled to a few degrees kelvin and comparing spectra of left and right-handed molecules using lasers calibrated to some of the world’s best primary frequency standards. These measurement will be hundreds of times more sensitive than previous experiments~\cite{DarStoZri10,tokunaga_probing_2013,Cournol_2019}. We have already developed the methods needed to bring organometallic species into the gas-phase and cool them down to a few kelvins in a cryogenic buffer-gas cell~\cite{Tokunaga2017,Asselin2017,darquie_valence-shell_2021,FieHaaSal22}. We will conduct Ramsey spectroscopy~\cite{shelkovnikov_stability_2008} on a slow beam of molecules extracted from the cold cell using frequency-stabilised lasers. We will alternate measurements between left- and right-handed molecules, thus performing a differential measurement where many systematic effects cancel out. We expect to achieve an uncertainty of $10^{-15}$ on the relative frequency difference between the enantiomers. The required frequency range, which is in the mid-infrared, is a difficult frequency band for lasers. To solve this problem, we have developed highly-stable quantum cascade lasers (QCLs)~\cite{argence_quantum_2015,santagata_high-precision_2019,tran_near-_2024,tran_extending_2025} and CO$_ 2$ lasers~\cite{chanteau_mid-infrared_2013} calibrated via the French research infrastructure REFIMEVE~\cite{cantin_accurate_2021} to Cs SI standard of time fountain clocks of the national metrology institute. We have validated this technology in the 1000~cm$^{-1}$ region and our current objective is to demonstrate access to the entire 500--2000~cm$^{-1}$ spectral window where QCLs are more or less readily commercially available~\cite{osmocene}. We have already at our disposal a 1600~cm$^{-1}$ source and the world’s first $\sim600$~cm$^{-1}$ continuous-wave distributed feedback QCL~\cite{wang_wavelength_2025,manceau_demonstration_nodate}.

\section{Results and discussion}

\subsection{Characterization and comparison to theory}

\begin{figure}[ht!]
\begin{center}
\includegraphics[scale=0.6]{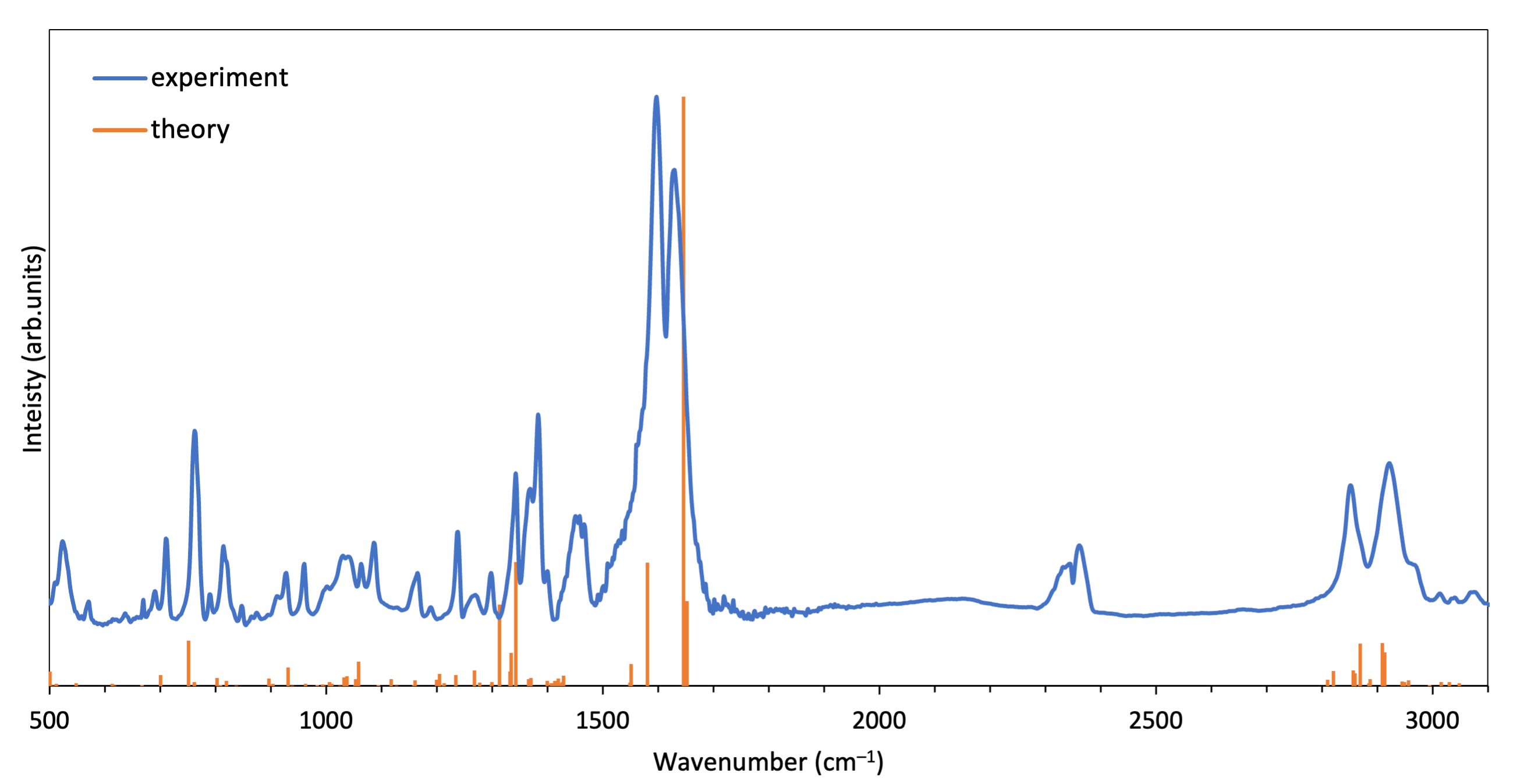}
\caption{Comparison of the experimental (KBr matrix) and calculated IR spectra of Pt(Me-BPCH). Both spectra are normalized to the highest peak. The calculated stick spectrum is based on harmonic vibrational analysis with empirically scaled frequencies using a factor 0.94. The experimental spectrum is vertically shifted for increased legibility.}
\label{fig:spectra}
\end{center}
\end{figure}

Pt(Me-BPCH) has been characterized in the solid state by infrared (IR) spectroscopy and X-ray diffraction. It is readily soluble in organic solvents and the NMR spectra are consistent with the presence of a single stereoisomer. Thus the solution structure is most likely the same as the solid state structure, and this structural robustness is confirmed by its homogeneity and yield following repeated recrystallization from solution. 
The calculated and experimental XRD structures are almost superimposable (Figure S14 in Supplementary Material) with RMSD of 0.33~{\AA}, or 0.1~{\AA} excluding hydrogen atoms.
The solid state IR matches well with the calculated vibrational modes in both energies and intensities of the key bands, Figure~\ref{fig:spectra}. The most striking feature of the IR spectrum are the two intense peaks at 1629~cm$^{-1}$ and 1597~cm$^{-1}$, corresponding to the amide carbonyl stretch and the pyridyl C--C stretch, respectively. Both of these asymmetric vibrations are associated with large changes in the permanent dipole moment in the direction perpendicular to the main $C_2$ molecular axis, which in turn causes the observed intense IR absorption.
Experimental details on XRD, NMR and IR measurements are given in the Supplementary Material.

\subsection{Sensitivity to Parity Violation}

The calculated frequencies and intensities of selected vibrational modes of experimental interest, along with the corresponding PV frequency shifts for Pt(Me-BPCH) and its derivatives Pt(CF$_3$-BPCH) and Au(Me-BPCH) are presented in Tables~\ref{tab:PtMe}--\ref{tab:PtCF3}.
These modes were selected based on their (i) frequencies in the experimentally accessible 500--2000~cm$^{-1}$ frequency range, (ii) measurable PV frequency shifts (at least one order of magnitude larger than the projected experimental sensitivity), and (iii) remarkably high intensities.
These selection criteria can be condensed to a simple quantitative metric $(\log I + |\Delta \nu^\text{PV}_{0\rightarrow 1}|)$, where $I$ is intensity in km/mol and $\Delta \nu^\text{PV}_{0\rightarrow 1}$ is the PV shift in Hz. Based on this metric together with minimal requirements for both properties ($|\Delta \nu^\text{PV}_{0\rightarrow 1}|>0.3$~Hz, $I>0.5$~km/mol), we have selected the top 20 modes for each molecule presented in Tables~\ref{tab:PtMe}--\ref{tab:PtCF3}.
The full list of 86 investigated vibrational modes for each molecule in the range 500--2000~cm$^{-1}$ is given in Tables~S10--S12 in the Supplementary Material. 

\begin{table}[ht!]
\caption{Selected vibrational modes in Pt(Me-BPCH) promising for PV measurements together with their respective calculated harmonic frequencies $\nu$ (unscaled), intensities $I$, $C_2$ symmetry irreps, and PV shifts $\Delta  \nu^\text{PV}_{0\rightarrow 1}$. For vibrational mode assignment, we use the usual simplified spectroscopic notation: $\nu$ stretching, $\delta$ (in-plane) bending, $\gamma$ out-of-plane bending, $\Delta$ and $\Gamma$ denote ring/skeletal deformations, in-plane and out-of-plane, respectively.}
\label{tab:PtMe}
\centering
\begin{tabular}{cccclcc}
\hline        
Mode &  ${\nu}$ & $I$  &$C_2$& Vib. assignment & $\Delta  \nu^\text{PV}_{0\rightarrow 1}$   &$\left|\Delta \nu^\text{PV}_{0\rightarrow 1}/\nu\right|$\\
  & (cm$^\text{--1}$)& (km/mol) &sym. & & (Hz)  & ($10^{-14}$)\\\hline
 34&  530.7&  3.2&  A&  $\Gamma$(skeletal)&  1.52& 9.53\\
 39&  582.7&  4.5&  B&  $\Delta$(skeletal)&  --1.52& 8.72\\
 40&  652.5&  3.5&  A&  $\Delta$(skeletal)&  --3.45& 17.65\\
 45&  745.0&  16.9&  B&  $\Gamma$(skeletal)&  1.61& 7.20\\
 47&  799.3&  70.5&  B&  $\gamma_\text{CH}$(pyr)&  0.42& 1.74\\
 49&  854.2&  12.1&  A&  $\gamma_\text{CH}$(pyr)&  0.90& 3.50\\
 50&  854.2&  12.6&  B&  $\gamma_\text{CH}$(pyr)&  0.91& 3.54\\
 76&  1126.5&  36.0&  A&  $\Delta$(c-hex)&  0.64& 1.88\\
% 88&  1313.4&  13.0&  B&  $\Delta$(pyr) + $\Delta$(PtNCCN)&  --0.80& 2.03\\
 89&  1313.6&  17.8&  A&  $\Delta$(pyr) + $\Delta$(PtNCCN)&  0.73& 1.84\\
 96& 1396.5& 123.6& B& $\delta_\text{CH}$(c-hex)& 1.18&2.83\\
% 97& 1411.9& 0.1& A& $\delta_\text{CH}$(c-hex)& 0.71&1.67\\
 99& 1418.9& 42.0& A& $\delta_\text{CH}$(Me) + $\Delta$(PtNCCN)& 0.39&0.92\\
 100& 1427.2& 53.3& B& $\delta_\text{CH}$(Me) + $\Delta$(PtNCCN)& 1.21&2.84\\
 101& 1428.9& 204.3& A& $\delta_\text{CH}$(Me) + $\Delta$(PtNCCN)& 1.21&2.83\\
 103& 1458.8& 13.9& A& $\nu_\text{CN}$(pyr) + $\delta_\text{CH}$(Me/pyr)& 2.12&4.85\\
 109& 1508.4& 14.8& A& $\delta_\text{CH}$(Me)& 1.00&2.21\\
 112& 1520.6& 13.5& B& $\nu_\text{CC}$(pyr) + $\delta_\text{CH}$(pyr)& 0.84&1.84\\
 115& 1649.9& 34.5& A& $\nu_\text{CN}$(pyr) + $\delta_\text{CH}$(pyr)& 0.54&1.10\\
 116& 1681.8& 193.3& B& $\nu_\text{CC}$(pyr) + $\delta_\text{CH}$(pyr)& 0.68&1.35\\
 118& 1751.6& 926.2& B& $\nu_\text{CO}$& --1.49&2.84\\
 119& 1757.8& 133.8& A& $\nu_\text{CO}$& 0.97&1.84\\
 \hline
\end{tabular}
\end{table}

\begin{table}[ht!]
\caption{Selected vibrational modes in Au(Me-BPCH) promising for PV measurements. For notation, see caption of Table 1.}
\label{tab:AuMe}
\centering
\begin{tabular}{cccclcc}
\hline        
Mode &  ${\nu}$ & $I$  &$C_2$& Vib. assignment & $\Delta  \nu^\text{PV}_{0\rightarrow 1}$   &$\left|\Delta \nu^\text{PV}_{0\rightarrow 1}/\nu\right|$\\
  & (cm$^\text{--1}$)& (km/mol) &sym. & & (Hz)  & ($10^{-14}$)\\\hline
 34&  512.7&  20.2&  B&  $\Gamma$(skeletal)&  --0.79& 5.13\\
 35&  518.0&  0.9&  A&  $\Gamma$(skeletal)&  2.93& 18.85\\
 40&  636.2&  2.0&  A&  $\Delta$(skeletal)&  2.10& 11.03\\
 42&  694.2&  21.2&  B&  $\Delta$(skeletal)&  2.66& 12.77\\
 60&  963.9&  28.5&  A&  $\nu_\text{PtN}$ + $\delta_\text{CH}$&  0.82& 2.84\\
 63&  1020.5&  12.0&  B&  $\delta_\text{CH}$(Me)&  0.94& 3.07\\
 64&  1039.1&  24.3&  B&  $\Delta$(pyr)&  3.45& 11.09\\
 65&  1044.3&  3.1&  A&  $\Delta$(pyr)&  2.19& 7.01\\
 73&  1101.1&  27.3&  B&  $\Delta$(c-hex)&  --1.01& 3.05\\
% 88& 1311.0& 15.3& B& $\Delta$(pyr) + $\Delta$(AuNCCN)& --0.79&2.01\\
 89& 1311.8& 24.9& A& $\Delta$(pyr) + $\Delta$(AuNCCN)& --1.48&3.77\\
 91& 1347.9& 21.3& B& $\delta_\text{CH}$(c-hex)& 0.69&1.70\\
% 96& 1397.2& 148.4& B& $\delta_\text{CH}$(c-hex)& --0.24&0.58\\
% 97& 1411.4& 0.0& A& $\delta_\text{CH}$(c-hex)& 0.59&1.39\\
 98& 1414.4& 136.8& B& $\delta_\text{CH}$(Me) + $\Delta$(AuNCCN)& --1.94&4.58\\
 99& 1419.4& 80.6& A& $\delta_\text{CH}$(Me) + $\Delta$(AuNCCN)& 0.43&1.01\\
% 100& 1428.0& 210.1& A& $\delta_\text{CH}$(Me) + $\Delta$(PtNCCN)& 0.23&0.53\\
 102& 1454.1& 62.6& B& $\nu_\text{CN}$(pyr) + $\delta_\text{CH}$(Me/pyr)& 2.66&6.11\\
 103& 1459.3& 32.4& A& $\nu_\text{CN}$(pyr) + $\delta_\text{CH}$(Me/pyr)& 1.50&3.43\\
 112& 1516.0& 32.2& A& $\nu_\text{CC}$(pyr) + $\delta_\text{CH}$(pyr)& 0.51& 1.12\\
 115& 1654.9& 62.3& A& $\nu_\text{CN}$(pyr) + $\delta_\text{CH}$(pyr)& 0.63&1.27\\
 116& 1672.7& 282.6& B& $\nu_\text{CC}$(pyr) + $\delta_\text{CH}$(pyr)& 0.91&1.82\\
 118& 1730.9& 858.9& B& $\nu_\text{CO}$& 1.16&2.23\\
 119& 1741.7& 76.7& A& $\nu_\text{CO}$& 0.81&1.55\\
 \hline
\end{tabular}
\end{table}

\begin{table}[ht!]
\caption{Selected vibrational modes in Pt(CF$_3$-BPCH) promising for PV measurements. For notation, see caption of Table 1.}
\label{tab:PtCF3}
\centering
\begin{tabular}{cccclcc}
\hline        
Mode &  ${\nu}$ & $I$  &$C_2$& Vib. assignment & $\Delta  \nu^\text{PV}_{0\rightarrow 1}$   &$\left|\Delta \nu^\text{PV}_{0\rightarrow 1}/\nu\right|$\\
  & (cm$^\text{--1}$)& (km/mol) &sym. & & (Hz)  & ($10^{-14}$)\\\hline
 42&  541.9&  13.0&  B&  $\Delta$(skeletal)&  --1.47& 9.04\\
 48&  656.6&  3.1&  A&  $\Delta$(skeletal)&  --2.20& 11.18\\
 55&  774.8&  10.4&  B&  $\Gamma$(skeletal)&  1.94& 8.35\\
 57&  797.0&  69.9&  B&  $\gamma_\text{CH}$(pyr)&  0.88& 3.69\\
 62&  873.9&  27.4&  B&  $\gamma_\text{CH}$(pyr)&  0.82& 3.12\\
 70&  987.1&  10.2&  A&  $\nu_\text{PtN}$ + $\delta_\text{CH}$&  0.87& 2.93\\
 78&  1098.2&  22.8&  A&  $\nu_\text{CC}$(c-hex)&  0.61& 1.85\\
 82&  1126.5&  60.2&  A&  $\Delta$(c-hex)&  0.50& 1.48\\
 83&  1146.6&  35.9&  A&  $\nu_\text{CF}$&  0.95& 2.77\\
% 85&  1164.2&  412.8&  B&  $\nu_\text{CF}$&  0.05& 0.14\\
% 92& 1230.2& 204.3& A& $\nu_\text{CF}$& 0.04&0.11\\
% 98& 1310.4& 66.2& A& $\Delta$(pyr) + $\Delta$(PtNCCN)& 0.18&0.45\\
 99& 1310.6& 12.9& B& $\Delta$(pyr) + $\Delta$(PtNCCN)& --1.14&2.90\\
 102& 1351.0& 104.2& A& $\nu_\text{CC}$(CF3) + $\delta_\text{CH}$& 0.36&0.90\\
 103& 1352.4& 111.7& B& $\nu_\text{CC}$(CF3) + $\delta_\text{CH}$& 0.39&0.96\\
 108& 1401.8& 169.1& B& $\delta_\text{CH}$(c-hex)& 0.87&2.08\\
% 109& 1415.0& 3.1& A& $\delta_\text{CH}$(c-hex)& 0.52&1.22\\
 110& 1428.7& 187.0& B& $\Delta$(PtNCCN)& 2.23&5.21\\
 111& 1431.8& 205.8& A& $\Delta$(PtNCCN)& 1.80&4.20\\
 112& 1479.6& 3.5& B& $\nu_\text{CN}$(pyr) + $\nu_\text{CC}$(CCO)& 1.74&3.92\\
 113& 1480.5& 6.3& A& $\nu_\text{CN}$(pyr) + $\nu_\text{CC}$(CCO)& 3.24&7.31\\
 122& 1680.9& 127.9& B& $\nu_\text{CC}$(pyr) + $\delta_\text{CH}$(pyr)& 0.61&1.21\\
 124& 1755.9& 927.0& B& $\nu_\text{CO}$& --1.47&2.80\\
 125& 1762.2& 121.0& A& $\nu_\text{CO}$& 1.20&2.28\\
 \hline
\end{tabular}
\end{table}

%Brief overall description
The selected modes exhibit relative PV sensitivities in the $10^{-14}-10^{-13}$ range in all three molecules. 
In Pt(Me-BPCH), the largest PV shifts are observed in the skeletal deformation (both in-plane and out-of-plane) modes 34, 39, 40 and 45, the pyridine C--N stretching mode 103, and the asymmetric C=O stretching mode 118. This is not surprising, as these modes exhibit either big collective movement (modes 34, 39, 40, 45) or significant changes in the charge distribution (modes 103, 118). Significant changes in the electronic density around the heavy atom (Pt) are tied to stronger PV sensitivity  due to associated larger changes in the chiral density around the heavy atom.

Contrary to our expectation, the Au substitution did not lead to the anticipated overall increase in PV sensitivity. Instead, it lead to redistribution of the PV response over different modes while keeping the overall sensitivity roughly equal in magnitude. Nevertheless, also in case of Au(Me-BPCH), we observe large PV shifts for some of the skeletal deformation modes (35, 40, 42) and also in the pyridine C--N stretching modes 102 and 103, and the asymmetric C=O stretching mode 118, reminiscent of the platinum complex. In addition for Au(Me-BPCH), we observe large shifts in the in-plane pyridine and AuNCCN ring deformation modes 64, 65 and 98. These modes do have analogs also in Pt(Me-BPCH) (modes 66, 67 and 98), but these are weaker both in therms of the PV sensitivity and in terms of the IR absorption intensity.

Finally, in Pt(CF$_3$-BPCH), the most PV sensitive modes are also of the discussed types -- skeletal deformations (42, 48, 55), in-plane pyridine and PtNCCN ring deformations (99, 110, 111), pyridyl C--N stretches (112, 113) and the C=O stretch (124). In contrast to the the other two systems lacking fluorine atoms, here, we also observe a fairly strong shift in the C--F stretching mode 83 which falls into the spectral window of interest, unlike the associated high-frequency methyl C--H stretches in Pt(Me-BPCH) and Au(Me-BPCH).

%Comparing vibrational structures
As the three investigated molecules are closely related and structurally very similar, one can expect also their vibrational structures (fingerprints) to resemble one another. This is also clear from a quick glance at the vibrational assignments in Tables~\ref{tab:PtMe}--\ref{tab:PtCF3} and the discussion above.
Especially, in the case of Pt(Me-BPCH) and Au(Me-BPCH), where not only the bonding patterns are equivalent but also all the atomic masses are near-identical, the vibrational modes follow a very closely related pattern.
On the other hand, in Pt(CF$_3$-BPCH), the hexa-fluorination shifts all modes previously associated with methyl C--H stretching and bending towards lower frequencies and allows these then to couple to other modes in the similar frequency range. This leads to a lower correspondence between vibrational structures of Pt(Me-BPCH) and Pt(CF$_3$-BPCH).
To better quantify the correspondence between the modes of the three molecules, we have calculated the overlaps (normalized dot products) between the displacement vectors (Table~S13 in Supplementary Material).
Based on these, we find that out of the 86 modes in the laser window (500--2000~cm$^{-1}$), 79 modes match closely between Pt(Me-BPCH) and Au(Me-BPCH), while only 53 modes match between Pt(Me-BPCH) and Pt(CF$_3$-BPCH).
Especially in the frequency ranges 750--970~cm$^{-1}$ (modes 46--59), 1070--1415~cm$^{-1}$ (modes 70--97), and 1495--1760~cm$^{-1}$ (modes 106--119), we find almost an one-to-one correspondence between the modes of all three molecules.

Even though the vibrational structures resemble each other to a high degree, the PV response of the corresponding modes between different molecules can be markedly different. Perhaps, the most prominent case is the aforementioned asymmetric C=O stretching mode (mode 118 in Pt(Me-BPCH) and Au(Me-BPCH) and mode 124 in Pt(CF$_3$-BPCH)). Although this mode is highly IR intense and PV sensitive in all three systems, its sign is different in the Au-complex compared to both Pt-complexes. This is not the case, however, for the symmetric C=O stretching mode, where the sizes and signs of the PV shift in the three molecules agree.
Another such example is the in-plane skeletal deformation mode 40 in both $M$(Me-BPCH) ($M=$~Pt, Au) and mode 48 in Pt(CF$_3$-BPCH). Again, the gold complex is the odd one out with the opposite sign of the PV shift. Altogether, we identified 16 such discrepancies between the PV shifts of the related modes in Au-complex compared to the Pt-complexes (Tables~S10--S13 in the Supplementary Material). On the other hand, not a single case of such sign discrepancy is observed between Pt(Me-BPCH) and Pt(CF$_3$-BPCH). The change in electronic structure between closed-shell Pt-complexes and open-shell Au-complex results in these PV sensitivity redistributions. Based on this observation, it is likely a good strategy to explore multiple derivatives of organometallic complexes before selecting the final measurement target, as these often share many other physical and chemical properties while offering a wider variety of PV responses.

%A/B - odd/even / sym pairs
Throughout the discussion so far, we have loosely used the terms symmetric and asymmetric modes, more specifically referring to $C_2$ point group irreducible representations (irreps) A and B.
These symmetries have direct important consequences for the spectroscopically relevant IR intensities and PV sensitivities of the different vibrational modes.
To aid the discussion, we quantified the matching of symmetric/asymmetric (A/B) mode pair in terms of overlaps between the displacement vectors of A modes and the symmetrized displacement vectors of the B modes and found such symmetry pairs spanning the whole investigated range (Table~S14 in Supplementary Material). 

For example, in Pt(Me-BPCH), the low-frequency modes 39, 45 and 47, but also the high-frequency mode 116 shown in Table~\ref{tab:PtMe} all belong to the asymmetric irrep B. Their corresponding symmetric pairs (modes 38, 44, 46 and 117) while exhibiting equivalent PV shifts all have near zero intensities.
This is of course not a general conclusion, as can be clearly seen from symmetry pairs 49/50 or 100/101 in Table~\ref{tab:PtMe}, which have both similar PV shifts as well as intensities.
The distinction between the former and the latter examples lies in the character of these modes. Modes 38/39, 44/45, 46/47, 116/117 represent vibrations that induce dipole moment changes perpendicular to the molecular $C_2$ axis, which add up in B modes, but cancel in the A modes. In contrast, modes 49/50 and 100/101 induce dipole moment changes parallel to the molecular axis, thus no intensity suppression occurs. 
%A: 89 +PV  (B: 88 -PV)

% PT formula
The PV vibrational shifts can be analyzed further using the analytic expression previously derived by means of perturbation theory~\cite{wormit2014strong,schwerdtfeger2005relativistic}:
\begin{equation} \label{eq:pertexp}
\Delta\nu^\text{PV}_{0\rightarrow 1}\approx\Delta\nu^\text{PV}_\text{quad}+\Delta\nu^\text{PV}_\text{anharm}=\frac{P^{[2]}}{2\pi\mu\omega_e}-\frac{P^{[1]}V^{[3]}}{2\pi\mu^2\omega_e^3}
\end{equation}
where $\mu$ and $\omega_e$ are the reduced mass and the harmonic vibrational frequency of the normal mode under consideration, and $V^{[k]}$ and $P^{[k]}$ are the $k$-th derivatives at equilibrium geometry of the corresponding potential energy curve $V$ and the PV potential curve $V^\text{PV}$, respectively. 
There are two contributing terms. We call the first term quadratic as it is proportional to the second derivative of the $V^\text{PV}$ curve, and we call the second anharmonic because it is proportional to the product of the first derivative of the $V^\text{PV}$ curve and the third derivative of the $V$ curve (anharmonicity). The values of these contributions are given for all investigated modes in Tables~S10--S12 in the Supplementary Material.

Due to symmetry restrictions on the atomic displacement, all asymmetric (irrep~B) vibrational modes necessarily exhibit even potential energy functions $V$ as well as even PV potential functions $V^\text{PV}$ (i.e. they can be expanded only in terms of even derivatives). No such restriction applies, however, to the symmetric (irrep~A) vibrational modes, for which both $V$ and $V^\text{PV}$ curves can in principle take on any shape. Nevertheless, most irrep-A modes have fairly harmonic $V$ curves, except for the stretching modes, where anharmonicity can be notable. This in turn means that for all B modes, the anharmonic contribution $\Delta\nu^\text{PV}_\text{anharm}$ is strictly zero and only the quadratic term $\Delta\nu^\text{PV}_\text{quad}$ survives. For the A modes, sufficient anharmonicity in the $V(q)$ curves is required to observe $\Delta\nu^\text{PV}_\text{anharm}$ contributions of appreciable size. Only a handful of modes in each molecule exhibit values $|\Delta\nu^\text{PV}_\text{anharm}| > 0.3$~Hz.

\begin{figure}
\begin{center}
\tiny{Pt(Me-BPCH), mode 118 \hspace{1.65cm} Au(Me-BPCH), mode 118 \hspace{1.65cm} Pt(CF$_3$-BPCH), mode 124}
\includegraphics[scale=0.28,trim=0 0 0 2mm,clip]{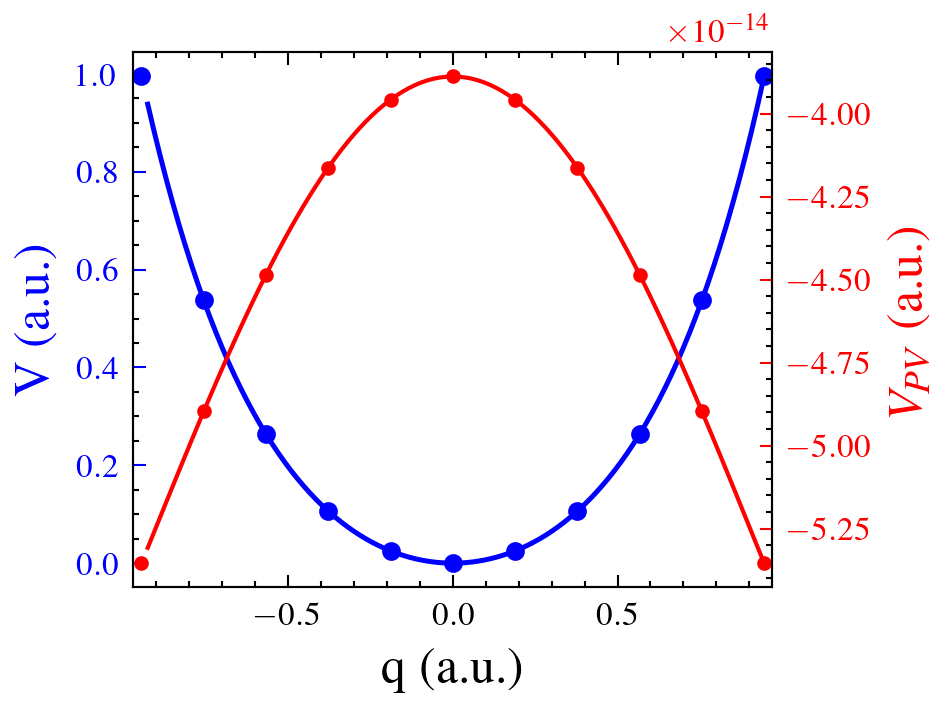}
\includegraphics[scale=0.28,trim=0 0 0 2mm,clip]{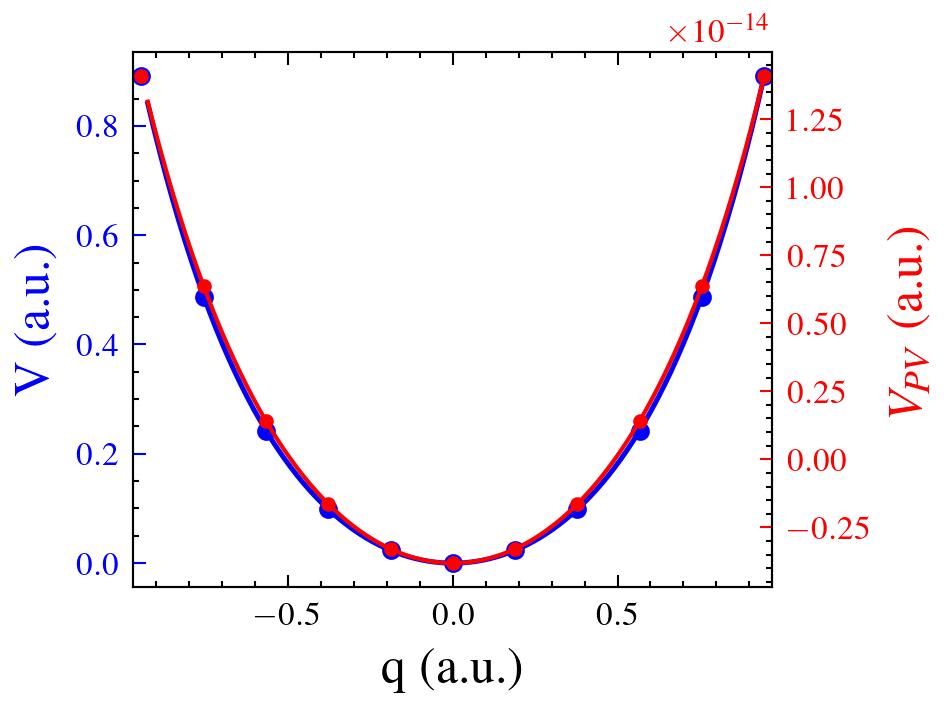}
\includegraphics[scale=0.28,trim=0 0 0 2mm,clip]{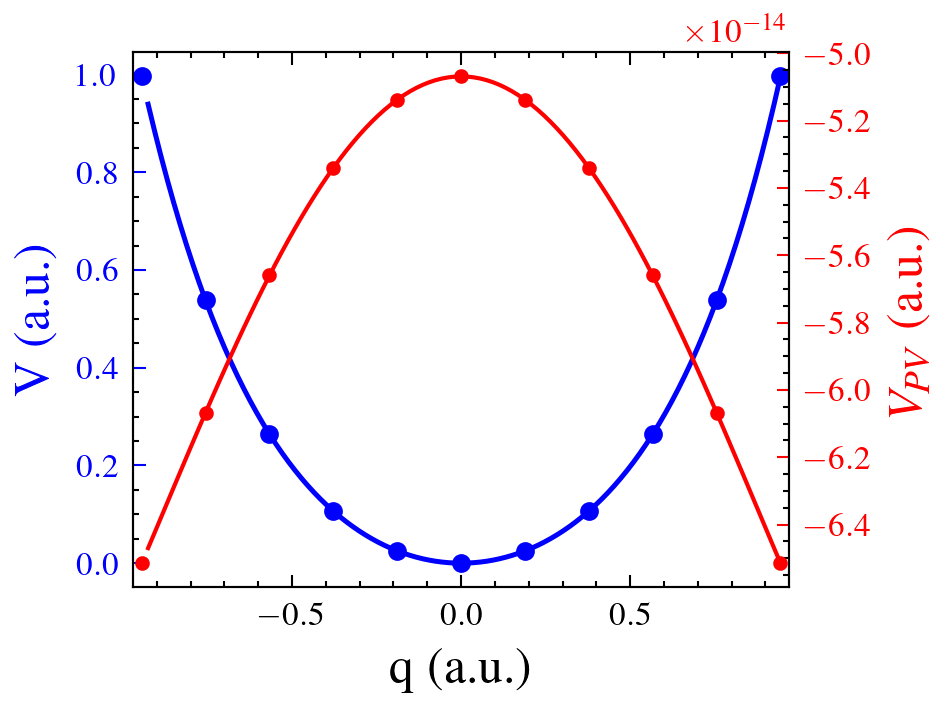}
\tiny{Pt(Me-BPCH), mode 119 \hspace{1.65cm} Au(Me-BPCH), mode 119 \hspace{1.65cm} Pt(CF$_3$-BPCH), mode 125}
\includegraphics[scale=0.28,trim=0 0 0 2mm,clip]{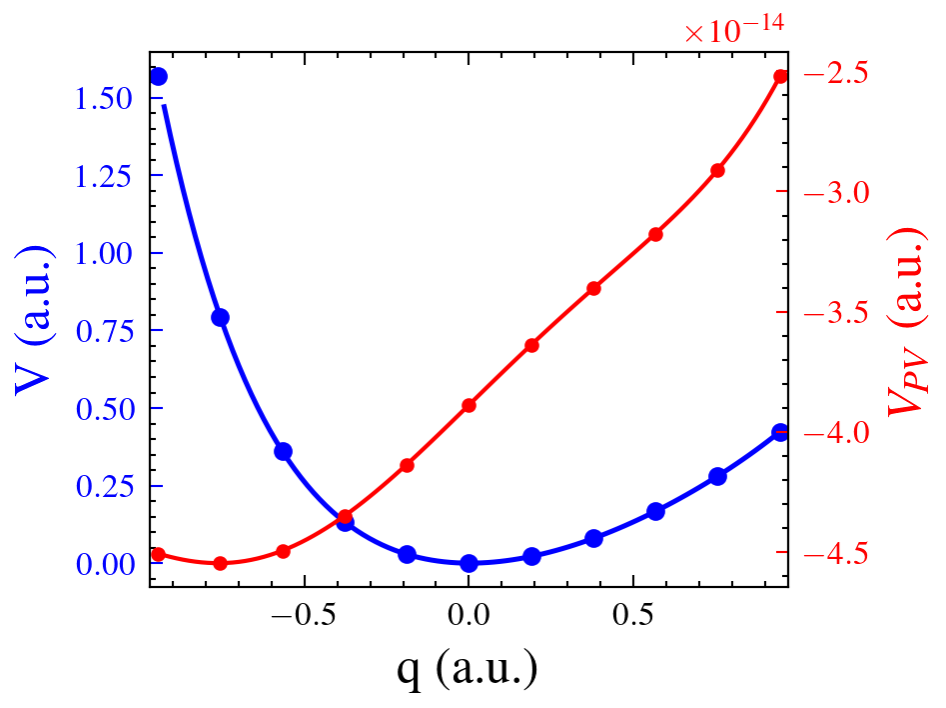}
\includegraphics[scale=0.28,trim=0 0 0 2mm,clip]{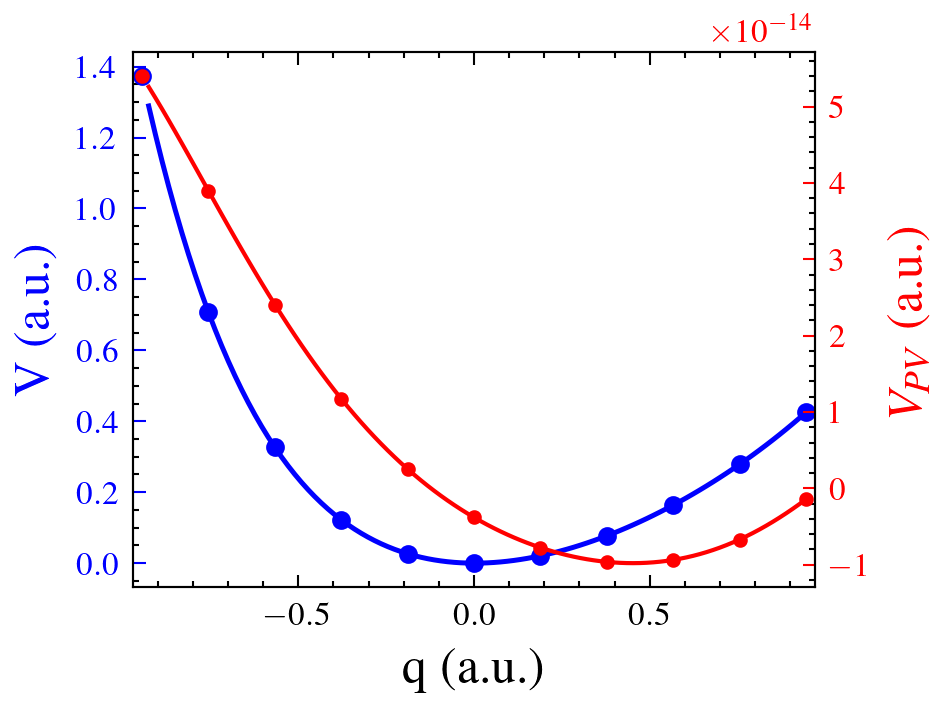}
\includegraphics[scale=0.28,trim=0 0 0 2mm,clip]{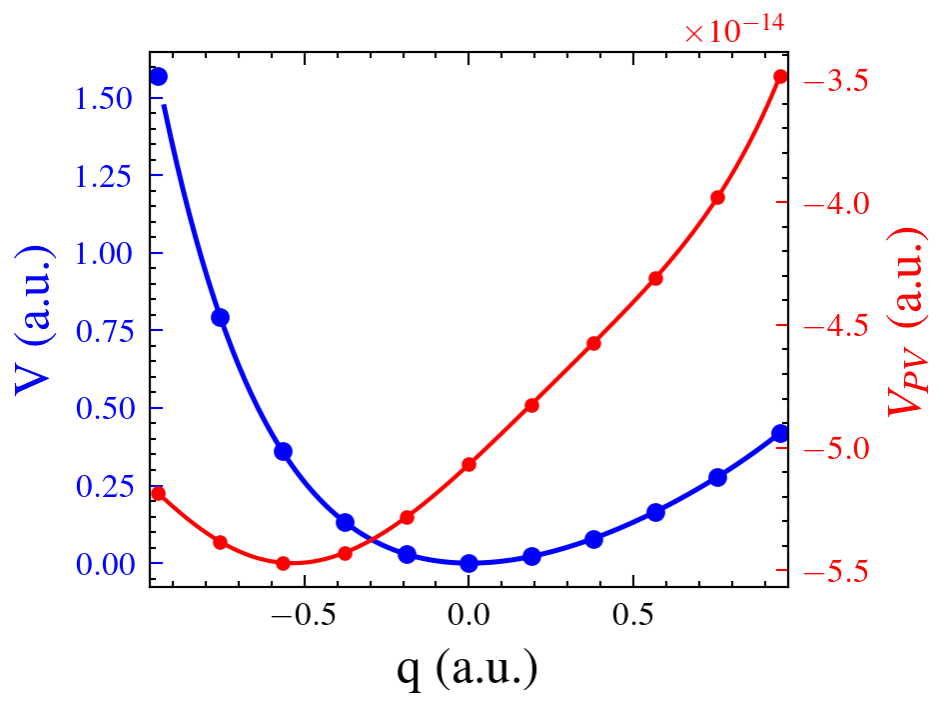}
\caption{Comparison of the potential energy curves $V(q)$ (blue lines) and the PV potential curves $V^\text{PV}(q)$ (red lines) of the asymmetric (top) and symmetric (bottom) C=O strecthing modes of Pt(Me-BPCH), Au(Me-BPCH) and Pt(CF$_3$-BPCH). The displacement vectors (i.e. the $q$ axes) are aligned and oriented in the same direction for easier visual comparison.}
\label{fig:modes}
\end{center}
\end{figure}

We illustrate the discussion above with an example of the carbonyl stretching modes in Figure~\ref{fig:modes}. Both symmetric and asymmetric C=O stretching modes are present in all three investigated molecules with a near perfect overlap of the corresponding displacement vectors as well as near perfect A/B correspondence (Tables~S13--S14 in the Supplementary Material).
In Figure~\ref{fig:modes}, one can clearly see that for the asymmetric modes (top plots), both $V(q)$ and $V^\text{PV}(q)$ functions are indeed even. On the other hand, the symmetric modes (bottom plots) are significantly anharmonic and the corresponding $V^\text{PV}(q)$ functions contain both odd and even contributions. These are some of the very few modes where the $\Delta\nu^\text{PV}_\text{anharm}$ term dominates.

Furthermore, Figure~\ref{fig:modes} also illustrates well the significant difference between the PV responses of the Au-complex (middle) and the Pt-complexes (left, right). In the asymmetric mode 118, the PV curve is vertically flipped in Au(Me-BPCH) and so the second derivative and hence also the $\Delta\nu^\text{PV}_{0\rightarrow 1}$ change sign compared to the Pt-complexes. In the symmetric mode 119 of Au(Me-BPCH), the $V^\text{PV}(q)$ curve is also markedly different compared to the Pt-complexes. While in Pt(Me-BPCH) and Pt(CF$_3$-BPCH) the $V^\text{PV}(q)$ curves show only minimal curvature in the equilibrium region, and thus the corresponding quadratic contribution $\Delta\nu^\text{PV}_\text{quad} \approx 0$, in Au(Me-BPCH) the quadratic term dominates. The anharmonic contribution $\Delta\nu^\text{PV}_\text{anharm}$ is large for all three systems, however of opposite sign following the slopes of $V^\text{PV}(q)$ shown in Figure~\ref{fig:modes}.
In addition to Figure~\ref{fig:modes}, we provide other examples of analogous plots comparing relevant modes (and mode pairs) in Table~S15 in the Supplementary Material. These provide a clear visual representation of the differences in the PV shifts observed in Tables~\ref{tab:PtMe}--\ref{tab:PtCF3}.

\subsection{Measurement prospects}

For the three species considered in this work, almost all transitions (93\%) falling in the 500--2000~cm$^{-1}$ range (Tables~S10--S12 in the Supplementary Material) show relative PV frequency shifts $|\Delta \nu^\text{PV}_{0\rightarrow 1}/\nu|$ larger than $10^{-15}$, the projected sensitivity aimed for in the LPL experiment, making them particularly attractive from the experimental perspective. Normal modes 60 and 76 of Pt(Me-BPCH), 78, 82 and 83 of Pt(CF$_3$-BPCH)  and 60, 64 and 73 of Au(Me-BPCH) are all in the laser window with currently available metrology-grade sources at LPL and all exhibit high intensities. Vibrational mode 64 of Au(Me-BPCH) looks especially promising given its very large  ($\sim10^{-13}$) predicted PV shift, exceeding by up to two orders of magnitude the projected instrumental sensitivity. This is on the same level as the most sensitive transition of helical osmocene found in our previous study~\cite{osmocene} in the 1000~cm$^{-1}$ region where proper laser technologies are readily available, but the transition in Au(Me-BPCH) is about an order of magnitude more intense. Tables~\ref{tab:PtMe}--\ref{tab:PtCF3} also list a series of vibrational modes exhibiting PV shifts ranging from $10^{-14}$ to above $10^{-13}$ and remarkable intensities as high as a few 100~km/mol in the 600~cm$^{-1}$ and 1600~cm$^{-1}$ spectral windows currently targeted at LPL. Importantly, Pt(Me-BPCH) possesses three modes 34, 39 and 40 exhibiting $\sim10^{-13}$ PV shifts in the 600~cm$^{-1}$ region currently investigated at LPL~\cite{wang_wavelength_2025,manceau_demonstration_nodate}. Among the various compounds synthesized and currently at disposal for measurements, these are the highest sensitivities to PV accessible with current laser technology. Finally, we observe in all three compounds remarkably active C=O stretching modes around 1750~cm$^{-1}$. The asymmetric stretches are in particular exceptionally intense, with transition strengths of about 900~km/mol. This, together with their predicted relative PV shifts exceeding $10^{-14}$ and the commercial availability QCLs in the corresponding spectral window, make these transitions very promising from an experimental perspective.

\subsection{Conclusions and outlook}

In summary, we have presented a synthetic route and characterization of a new organometallic complex Pt(Me-BPCH) and explored its suitability for a future PV detection experiment by means of electronic structure calculations predicting PV shifts in the vibrational spectra.
Many of these are of significant practical interest as they offer strong PV response, high intensity and wavenumbers in the available metrology-grade laser window.
Furthermore, we have compared Pt(Me-BPCH) to its two derivatives Au(Me-BPCH) and Pt(CF$_3$-BPCH) and analyzed connections between their vibrational structures and the respective PV sensitivities.
These derivatives offer additional variety in terms of selecting the best possible vibrational mode as the measurement target.
The next experimental step will be to investigate the stability upon heating of Pt(Me-BPCH) -- which is at disposal for measurements -- and how easily it can be brought into the gas phase (volatility). This study also motivates us to explore the proposed synthesis of the other two compounds, in particular the \text{CF$_3$}-substituted derivative Pt(CF$_3$-BPCH) which should exhibit an increased volatility~\cite{Fcarbon}, and to design and build at LPL a metrology-grade laser system in the 1750~cm$^{-1}$ spectral range, where QCLs are readily available, for future ultrahigh-resolution spectroscopic studies.

\section*{Acknowledgments}
The authors would like to thank Peter Schwerdtfeger for his friendship, generosity, guidance and mentorship, as well as many fruitful scientific discussions and collaborations.

The authors thank the Center for Information Technology of the University of Groningen for their support and for providing access to the H\'abr\'ok high-performance computing cluster. 
Eduardus wishes to acknowledge the Indonesia Endowment Fund for Education/\textit{Lembaga Pengelola Dana Pendidikan} \text{(LPDP)} for research funding. 
LFP acknowledges the support from the project number VI.C.212.016 of the talent programme VICI financed by the Dutch Research Council (NWO), and support from the Scientific Grant Agency of the Slovak Republic (project 1/0254/24).

\section*{Data Availability Statement}
Experimental NMR, IR and XRD data as well as additional calculated data referred to within the paper are collected in the Supplementary Material.
Crystallographic results have been deposited in the Cambridge Structural Database under the reference:
David Bohle CCDC 2419012, Experimental Crystal Structure Determination, 2025, DOI: 10.5517/ccdc.csd.cc2m65np

\section*{Conflict of interest}
The authors declare no conflict of interest.

\bibliographystyle{tfo}
\bibliography{references}
\end{document}